\documentclass[aps,pra,twocolumn,showpacs]{revtex4}
\usepackage{amsmath,amssymb,color, graphicx}

\begin{document}
\title{Secret Sharing with a Single $d$-level Quantum System}
\author{Armin Tavakoli$^1$, Isabelle Herbauts$^1$, Marek \.{Z}ukowski$^2$ and Mohamed Bourennane$^1$}
\affiliation{$^1$Department of Physics, Stockholm University,
SE-10691 Stockholm, Sweden\\
$^2$Institute of Theoretical Physics and Astrophysics, Uniwersytet Gda\'nski, PL-80-952 Gda\'nsk, Poland}


\date{\today}


\begin{abstract}
We give an example of a wide class of problems for  which quantum information protocols based on {\em multi-system} entanglement can be mapped into much simpler ones involving  one  system. 
Secret sharing is a cryptographic primitive which plays a central role in various secure multiparty computation tasks and management of keys in cryptography. In secret sharing protocols, a classical message is divided into shares given to recipient parties in such a way that some number of parties need to collaborate in order to reconstruct the message. Quantum protocols for the task commonly rely on  multi-partite GHZ entanglement. We present a multiparty secret sharing protocol which requires only sequential communication of a single quantum $d$-level system (for any prime $d$). It has huge advantages in scalabilility and can be realized with the state of the art technology. 
\end{abstract}



\maketitle


\section{Introduction}
Splitting a message into $N$ shares so that the original message can be reconstructed if and only if at least $k\leq N$ of the shares are known is called a $(N,k)$ secret sharing threshold scheme (the threshold is $k$). Secret sharing constitutes an important cryptographic primitive in protocols for secure multiparty computation including password-authenticated key agreement, hardware security modules, private querying of databases, and establishment of access codes with restricted access. The first secret sharing schemes were presented independently by Shamir and Blakley by means of classical algorithms to split the message and classical communication to distribute the shares \cite{S79, B79}.
 
In Shamir's $(N,k)$ secret sharing threshold scheme, the distributor chooses a set of $k$ positive integers, known only to him/her, $a_0,...,a_{k-1}\in\{1,...,P\}$, where  $P$ is some large prime.  The first integer $a_0$ is the secret. The $k$'th order polynomial  $p(x)=a_0+a_1x+\ldots+a_{k-1}x^{k-1}$, 
 is used for coding the data. For $l=1,...,N$ the distributor computes $p(l)$ and communicates the value only the $l$'th party.
If at least $k$ of the recipients collaborate, they can easily recover the secret $a_0$, whereas knowing fewer than $k$ shares yields no information on $a_0$. However, like many schemes in classical cryptography, Shamir's scheme is vulnerable to intercept-resend attacks on the communications of the distributor.


The security for cryptographic tasks can be enforced by introducing quantum resources \cite{BB84, GRTZ02}. Quantum methods for (classical) secret sharing by three parties in a form of cryptographic protocol based on three particle GHZ entanglement \cite{GHZ} were given in \cite{ZZWH98}. In an independent later development, secret sharing protocols for three or four parties were proposed in Ref. \cite{HBB99}. 
Secret sharing for arbitrary many parties exploiting multipartite qubit entanglement can be found in Ref. \cite{SSZ03}, wherein security issues were shown to be linked to Bell inequalities. A general secret sharing scheme using multipartite $d$-level entanglement is given in \cite{YLH08}. Also, general $(N,k)$ quantum secret sharing threshold schemes have been analyzed in Ref. \cite{CGL99}.

There are several experimental demonstrations of secret sharing schemes with quantum resources. Three and four partite secret sharing using entanglement were reported in Refs. \cite{CZ05,GKBW07}. However, entanglement-based protocols are not scalable. The difficulty of obtaining the required quantum correlations grows with the number of parties involved. 

Fortunately, a more scalable secret sharing (of classical data) can be achieved using only sequential communication of a single qubit, see Ref. \cite{STBKZW05}. The work reports a successful proof-of-principle experimental demonstration of six party secret sharing of such a kind. Nevertheless, the security of proposed secret sharing schemes is not as robust as the security of Quantum Key Distribution (QKD). This is discussed in Ref. \cite{QGWZ07,H07} for both the entanglement-based scheme of \cite{ZZWH98,HBB99} and the single qubit scheme of \cite{STBKZW05}.

In this letter, we present a $(N,N)$ secret sharing threshold scheme using a single $d$-level quantum system, 
for odd prime dimension $d$. We investigate eavesdropping attacks and security issues. Finally, we discuss the scalability and efficiency of our protocol in comparison to other schemes 
involving qudit systems. Our principal aim is to show that you can map GHZ state protocols extended to $d$-level systems into protocols involving sequential transfer of a single qudit (as this is a significant simplification of such schemes). We restrict $d$ to odd primes because our protocol uses a cyclic property of a set of Mutually Unbiased (orthonormal) Bases (MUBs). Many MUBs are still unknown \cite{DEBZ10}. Complete sets are only known for dimensions which are powers of prime numbers  \cite{WF89}. For this restricted set of dimensions, the algebraic property on which our scheme relies was found only  for odd prime dimensions. 

The relation of our single qudit scheme with respect to GHZ state qudit secret sharing can be thought to be similar to that of the BB84 QKD-protocol \cite{BB84} and the E91 QKD-protocol \cite{E91} based on entanglement. However, due to the in principle arbitrary number of parties involved, significant advantages of the single qudit scheme emerge with the growing number parties.


\section{Secret sharing using GHZ state correlations}
Let us first describe a secret sharing protocol using multipartite $d$-level entanglement, for which  $d$ is an odd prime. This particular protocol is outlined in \cite{YLH08}. 

The protocol is designed for $N+1$ party secret sharing and requires an $N+1$ partite $d$-level GHZ state:  
$|GHZ_d^{N+1}\rangle=\frac{1}{\sqrt{d}}\sum_{j=0}^{d-1}|j\rangle^{\otimes N+1}$. 
The party 1 ($R_1$) acting as the distributor prepares the GHZ state, keeps one particle, and distributes the remaining $N$ particles to the $N$ recipient parties. In the given run, each of the $N+1$ parties independently chooses one of $d$ possible bases in which the local particle is measured. 

For security purposes, all parties choose their measurement bases from a set of $d$ MUBs. 
 The unit vectors belonging to the full set of $d+1$ MUBs will be denoted as  $|e^{(j)}_l\rangle$ where $j=0,...,d$ labels the  basis and $l=0,...,d-1$ enumerates the vectors of the given basis. One has  for $j\neq j'$:
\begin{equation}\label{MUBcondition}
\left| \langle e^{(j)}_l| e^{(j')}_{l'}\rangle\right|^2=\frac{1}{d}.
\end{equation}
 Apart from the computational basis, for which we give the index $j=d$, and denote its states by $|k\rangle$, the remaining $d$ MUBs are given by
\begin{equation}\label{MUB}
|e^{(j)}_l\rangle=\frac{1}{\sqrt{d}}\sum_{k=0}^{d-1}\omega^{k(l+jk)}|k\rangle
\end{equation}
where $\omega=e^{2\pi i/d}$. It can be easily shown that \eqref{MUB} satisfies \eqref{MUBcondition} for all prime dimensions \cite{I81}. We will denote  by $M$  the set of all vectors belonging to the MUB defined by \eqref{MUB}, and its elements by $M_{l,j}$, with the meaning of the indices as above.

In each run of the experiment party $n$ (denoted by $R_n$) chooses randomly a measurement  basis $j_n$. The local measurement in the basis  projects his/her particle onto one of vectors $M_{l,j_n}$. This is governed by  the probability distribution   
\begin{eqnarray}
&P(l_1,\dots,l_{N+1}|j_1,...,j_{N+1})&\nonumber\\ 
&=\frac{1}{d^{N+2}}\left|\sum_{k=0}^{d-1}\omega^{-\sum_{n=1}^{N+1}(k l_n+k^2j_n)}\right|^2.&\nonumber \\ \nonumber
\end{eqnarray}
Perfect GHZ correlations are possible if 
\begin{equation}\label{QSSconditionGHZ}
\sum_{n=1}^{N+1} j_n=0 \mod{d}.
\end{equation}
In such a case,  only results satisfying $\sum_{n=1}^{N+1}l_n=0\mod{d}$ occur, and all sets satisfying this relation are equally probable. However, if  condition  \eqref{QSSconditionGHZ} does not hold, then the probability distribution of the results is uniform. This is easy to see once one realizes that \eqref{MUB} and \eqref{MUBcondition} implies that  $|\sum_{k=0}^{d-1}\omega^{k(l+jk)}|^2=d$  for $j\neq0$ and any $l$.

Once the measurements are performed, the parties announce their choices of $j_n$. The distributor checks condition \eqref{QSSconditionGHZ}. Only if it is satisfied, the round is treated as valid and is used for secret sharing. The local results satisfy $\sum_{n=1}^{N+1}l_n=0\mod{d}$, whereas a sum with one or more $l_n$ missing has an arbitrary value (mod $d$).
Thus even $N-1$ collaborating parties cannot learn the values obtained by the other two. But $N$ parties can establish the value of the remaining party.
As the choices of $j_n$ are random,  the protocol succeeds in $1/d$ of the cases. 


\section{Secret sharing with a single qudit}
Our protocol relies on a cyclic property of the set of MUBs:
there exist unitary transformations $U_{l'j'}$,  such that  for any $l',j'\in\{0,...,d-1\}$, 
 any  vector $M_{l,j} $ can be mapped into $M_{l+l',j+j'}$.  That is, elements of $M$ are mapped into elements of $M$.

Note that, for  any  vector  $M_{l,j}$ can be transformed into $M_{l+1,j}$ by applying the  transformation
\begin{equation}\label{X}
X_d=\sum_{n=0}^{d-1}\omega^n |n\rangle\langle n|
\end{equation}
Simply,  using \eqref{X} and \eqref{MUB}, one gets
\begin{eqnarray}\label{Xshift}
&X_d|e^{(j)}_l\rangle=\frac{1}{\sqrt{d}}\sum_{n=0}^{d-1}\omega^n|n\rangle\langle n| \sum_{k=0}^{d-1}\omega^{k(l+jk)}|k\rangle&\nonumber \\
&=\frac{1}{\sqrt{d}}\sum_{k=1}^{d-1} \omega^{k((l+1)+jk)}|k\rangle =|e^{(j)}_{l+1}\rangle. &
\end{eqnarray}
Also, any    $M_{l,j}$ can be transformed into $M_{l,j+1}$ by 
\begin{equation}\label{Y}
Y_d=\sum_{n=0}^{d-1} \omega^{n^2}|n\rangle\langle n|.
\end{equation}
This can be shown in a similar way. Thus,  by applying the  operator $U_{l'j'}=X_d^{l'}Y_d^{j'}$, 
any $M_{l,j}$ is mapped into $M_{l+l',j+j'}$. 

The  protocol  runs as follows.

($i$)  The distributor $R_1$, who by the nature of the task is always assumed to be an honest party, prepares the state $|e^{(0)}_0\rangle=\frac{1}{\sqrt{d}}\sum_{j=0}^{d-1} |j\rangle \in M$, which will be denoted by $|\psi_{0}^d \rangle$. 

($ii$)  $R_1$ picks  two random numbers $x_1,y_1\in\{0,...,d-1\}$, and performs on $|\psi_{0}^d\rangle$ the transformation  $X_d^{x_1}Y_d^{y_1}$. This gives  $|\psi_1^d \rangle \in M$. The state is sent to party $R_2$.

($iii$) For $n=2,...,N+1$, the party $R_n$ generates two independent random numbers $x_n,y_n\in\{0,...,d-1\}$,   and applies  $X_d^{x_n}Y_d^{y_n}$ to the  qudit $|\psi_{n-1}^d\rangle$ received from $R_{n-1}$. $R_n$'s action gives a state $|\psi_n^d\rangle$ which is sent to subsequent party $R_{n+1}$, except in case of $R_{N+1}$ who sends the qudit back to the distributor, $R_1$. 

($iv$) 
$R_{1}$ randomly chooses $J\in\{0,...,d-1\}$ and measures the qudit in the basis $\{|e_l^{(J)}\rangle\}_l$. The outcome is labeled $a\in\{0,...,d-1\}$.

($v$) In random order {\em only} parties $R_2,...,R_{N+1}$ announce their choice of $y_n$. The distributor  announces {\em only} 
whether the round is valid or not. It is valid provided:
\begin{equation}\label{QSScondition}
\sum_{n=1}^{N+1}y_n=J\mod{d}.
\end{equation} 
Otherwise the round is rejected. If the round is valid, the private data of the parties, $\{x_n\}$, satisfy globally 
\begin{equation}\label{correlation}
\sum_{n=1}^{N+1}x_n=a \mod{d}.
\end{equation} 
The data exhibit perfect correlations and thus can be used for secret sharing, as was the case for GHZ based protocols, provided $R_1$ resets his/her $x_1$ to $x_1^{(scrt)}=x_1-a$. Again, the probability of a valid round is $1/d$. 

($vi$) In order to check the security, for a randomly chosen (by the distributor) subset of the rounds, all parties  $R_2,...,R_{N+1}$  announce their values of their private data $x_n$ (in the same sequence as was the announcement of $y_n$'s). The distributor checks  \eqref{correlation}. If $R_1$ registers a substantial fraction check runs for which  \eqref{correlation} does not hold, $R_1$ declares the whole secret sharing attempt as corrupt
(more details on security checks later).

($vii$) If the secret sharing attempt is not corrupt, parties $R_2,...,R_{N+1}$, after exchanging all their data $x_n$ for a valid run, not used in the security check, can learn the otherwise secret value $x_1^{(scrt)}$, for the given run, earlier known only to the distributor $R_1$.

 
The protocol works because after  all the transformations the final state reads
\begin{multline}\label{finalstate}
|\psi_{final}^d\rangle=\left(\prod_{n=1}^{N+1}X_d^{x_n}Y_d^{y_n}\right) |\psi_{0}^d\rangle\\
=\frac{1}{\sqrt{d}}\left(|0\rangle+ \sum_{k=1}^{d-1}\omega^{\sum_{n=1}^{N+1}(kx_n+k^2y_n)}|k\rangle\right).
\end{multline} 
$R_{1}$'s measurement of \eqref{finalstate}  yields an outcome with unit probability, provided $|\psi_{final}^d\rangle$ is  an eigenstate of the measured observable. This happens if and only if \eqref{QSScondition} is satisfied. Otherwise,  $|\psi_{final}^d\rangle$ is some element $M_{l',j'}$ with $j'\neq J$ and thus by \eqref{MUBcondition} the probability of any outcome is $1/d$.

For a valid run the correlations are effectively equivalent to the ones for the GHZ based protocol: the choice of $y_n$ corresponds to  $R_n$'s choice of measurement basis, while $x_n$ is analogous to the local  outcome.


\section{Security discussion}

Protocols for secret sharing have to guarantee security. Consider an example of an attack by an external eavesdropper. If the eavesdropper, Eve, attempts an intercept-resend attack and intercepts the qudit, in the state $|e^{(j)}_l\rangle$, on the way form $R_k$  to $R_{k+1}$, she can choose one of $d$ relevant bases to measure. With probability $1/d$ she chooses a basis $j'=j$ and the attack succeeds, but with probability $\frac{d-1}{d}$ she has $j\neq j'$ in which case the state she sends to $R_{k+1}$ will be altered. The eavesdropping, to some extent depending on $d$, causes inconsistencies between the private data and condition \eqref{correlation}, and is therefore  detectable in step ($vi$) of the protocol. 


For more general eavesdropping attacks, in the qudit transfer from $R_k$ and $R_{k+1}$, we can regard the parties $R_1,...,R_k$ as a 'block' effectively representing a single party, and  parties $R_{k+1},...,R_{N+1}$ and $R_1$, acting as the measuring party, we can treat similarly. Thus, the attack is reducible to the scenario encountered in the BB84 two-party QKD (see e.g. \cite{CBKG02}) in which the sender and the receiver, both effectively our $R_1$, {\em do not} announce their bases, but only validity of a run. Generally, this makes security effectively perfect, even if Eve tries this strategy at more than one qudit transfer link.


An alternative  trick which can be used by Eve is to send via the unitary gate of partner $R_k$ one more qudit or even a multi qudit pulse, say separated in time, so that it can be somehow intercepted by her  beyond the gate, without intercepting the protocol qudit. After $y_k $ is announced she can learn the actual unitary transformation and thus $x_n$.
However, this is easily detectable, if $R_k$  makes the number of particles measurement at the exit of his/her gate (in some randomly chosen runs).

Yet another possibility is for Eve to intercept the qudit sent by $R_1$, and send a qudit of her own to $R_2$ in its stead.  Eve collects her qudit once it is sent by $R_{N+1}$, and waits for the announcement of $y_n$'s. The intercepted qudit of $R_1$ can be somehow manipulated by her, however it must reach the measurement station of  $R_1$ at the right time. 
After $y_n$'s are announced she can measure her qudit and recover the value $x_2+\ldots+x_{N+1} \mod{d}$. However, the attack will be detected in step ($vi$) of the protocol since $R_1$ performs the measurement before the $y_n$'s are announced. There is no way for Eve to perform a $y_n$'s dependent manipulation on a qudit which is already measured by $R_1$. 

\subsection{Discussing security against conspiracies}
In secret sharing one faces  the possibility of conspiring cheating subsets of parties. In the worst case, only the distributor $R_1$ and one more party are honest, leaving $N-1$ conspiring parties. Conspiracies significantly complicates the security analysis of secret sharing schemes and much is therefore unknown about security of various schemes. Here, we will discuss the robustness of our scheme against some particular conspiracies. However, rigorous security proof for general conspiracies is unknown.

In, e.g., Refs \cite{H07, HW10} eavesdropping attacks using quantum memories and entangling of systems with an ancilla were shown to lead to security problems in the protocol of Ref. \cite{STBKZW05}. However, the attacks of \cite{HW10} require that either the first or the final party are cheating, which never happens in our protocol as $R_1$ is effectively both first an last party. Additionally,  the eavesdropping attacks of  \cite{HW10} require knowledge of also $y_1$ and $J$, which is impossible since $R_1$ never announces any data.   

More generally, the cheaters could use some attack based on entangling the qudit with an ancilla, or possibly storing the protocol qudit in a quantum memory and creating a new entangled state, of which a subsystem is communicated further along the protocol loop. Still, they ought not to be able to profit. The reason is the absence of data announcement from $R_1$ renders the qudit available for the cheaters effectively in a mixed state, for which there is no observable which would give an outcome with  unit probability. Furthermore, if the cheaters combine their attack with eavesdropping, on the actions of the honest parties, they will be detected in step $(vi)$ of the protocol on basis of the arguments from the previous section. 

\section{Comparing secret sharing schemes}
There exists a number of quantum protocols for secret sharing. The protocols for three and four-partite secret sharing proposed in \cite{HBB99} and its generalization to high-level multipartite configurations \cite{YLH08} requires the preparation of a {\em high-fidelity} GHZ state with $N+1$ subsystems. With growing $N$ this becomes an increasingly difficult task. The experimental requirements make these schemes unscalable. Furthermore, another problem arises if we also consider inefficient detection. Let $\eta\in[0,1]$ be the detector efficiency. Given that the condition \eqref{QSSconditionGHZ} is satisfied for a particular round, it is required that all parties succeed with their measurements otherwise the round has to be rejected. The probability that all $N+1$ detection stations give a successful detection is $\eta^{N+1}$. Furthermore, note that in  GHZ state protocols $d (N+1)$ detectors are required. As each detection station introduces possible registration errors, the overall error would accumulate. However, such GHZ state protocols can enable security against device manipulation which is an important security feature when the experimenter does not fully control its own measuring device.

Consider now secret sharing with QKD involving qudits, in which the distributor uses $N$ pairwise independent QKD channels, each shared with one of the recipients. The protocol of such type which is directly comparable to our scheme involves encoding in $d$ different MUBs. For every round the distributor sends data $x_n$ to party $n$ such that suitable correlations are obtained to achieve secret sharing. However, using $d$-level QKD each recipient has a probability of $1/d$ to choose the correct basis. If the QKD scheme between the distributor and $R_n$ is repeated $m$ times, the probability that $R_n$ chooses the correct basis at least once is $1-\left(1-\frac{1}{d}\right)^m$. For successful secret sharing through QKD, the distributor has to repeat the scheme independently with each party until all of them report a correct choice of basis at least once. The probability, $p_{success}$, that for all $n=2,...,N+1$, $R_n$ has at least one correct choice is
$p_{success}=\left(1-\left(1-\frac{1}{d}\right)^m\right)^N$. Solving for the number of rounds, $m$, we find $
m=\left\lceil\frac{\ln\left(1-p_{success}^{1/N}\right)}{\ln\left(1-1/d\right)}\right\rceil$.
As an example, we can choose $N=10$, and pick $d$ large, say $d=23$, so that $p_{success}$ leads to a good estimate of the number of rounds required to distribute \textit{exactly} one number to each recipient. We require that the probability of success is somewhat high, say $p_{success}=0.8$. Then the approximate number of rounds required is about $m=86$. For distributing a secret of realistic size in many shares and to guaratee its security, one will typically need to distribute larger data sets. In this estimation we have not considered the parties having inefficient detectors.  Including this possibility decreases the protocol efficiency by an average factor of $\eta^{N}$. Therefore, such QKD-schemes requires much more rounds and detectors than our scheme. However, the security of QKD \cite{CBKG02} is more robust and well studied than that of secret sharing, which allows for higher security to the price of lower efficiency. 

The security can be further increased for such QKD schemes by performing the QKD in a device independent manner, i.e. with parties performing measurements on entangled state obtaining data that violates a Bell inequality. However, this also leads to an additional reduction of efficiency due to the low key rates high experimental requirements associated with device independent schemes.

In our protocol, for any $N$, only a single qudit is required. This enhances experimental feasibility: there is no issue of scalability of the initial state preparation. As  the protocol involves just one detector station  scalability is further enhanced. In addition, from the point of view of interferometry, our scheme is in the domain of single particle interference. It is well known that one can achieve very high interference visibilities in such cases whereas multiparticle interference effects for photons can accuire high visibilities only in the case of two qubits. Multiphoton qudit experiments will experience alignment problems, errors due to imperfections in the optical components and only partial distinguishability of photons comming from different sources \cite{P12}. 
For security purposes, it is very important to keep the quantum error rates to a minimum. However, our scheme requires control over the devices and the security against collective attacks remains unknown. Finally, we do note that our scheme requires the same number of local unitary operations as is used in the corresponding GHZ state protocol \cite{YLH08}, and is therefore subject to the same accumulation of noise due to imperfections in the local unitary actions.



 
\section{Conclusions} We have presented a secret sharing protocol using only a communication of a single qudit. While the security proofs are incomplete against possible sophisticated attacks which we did not include in our analysis, the scheme is secure against standard attacks. The scheme provides big advantages in scalability over earlier schemes and thus make proof-of-concept experiments feasible. Moreover, using our methods a wide class of quantum protocols using (multiparty) entanglement can be mapped into simple ones involving one qudit.




\begin{acknowledgments}
We thank Ingemar Bengtsson for sharing his knowledge about MUBs. The project was supported by the Swedish Research Council, ADOPT,  and QOLAPS project of ERC. MZ acknowledges the support of TEAM project of FNP.        
\end{acknowledgments}


\end{document}